\documentclass[a4paper,12 pt, openodd, a4paper]{article}

\usepackage{epsfig,amsmath,amssymb,graphics,color,calc,cite,psfrag}
\usepackage{amsmath}
\usepackage{amsthm}
\usepackage{amssymb}
\usepackage{amsopn}
\newtheorem{teo}{Theorem}[section]      \newtheorem{pro}[teo]{Proposition}
\newtheorem{defi}[teo]{Definition}      \newtheorem{lem}[teo]{Lemma}
\newtheorem{cor}[teo]{Corollary}        \newtheorem{rem}[teo]{Remark}
\newtheorem{con}[teo]{Condition}

\newcommand{\bteo}[1]{\begin{teo}\label{#1}}
\newcommand{\bpro}[1]{\begin{pro}\label{#1}}
\newcommand{\bdefi}[1]{\begin{defi}\label{#1}}
\newcommand{\blem}[1]{\begin{lem}\label{#1}}
\newcommand{\bcor}[1]{\begin{cor}\label{#1}}
\newcommand{\brem}[1]{\begin{rem}\label{#1}}
\newcommand{\bcon}[1]{\begin{con}\label{#1}}

\newcommand{\eteo}{\end{teo}}   \newcommand{\epro}{\end{pro}}
\newcommand{\edefi}{\end{defi}} \newcommand{\elem}{\end{lem}}
\newcommand{\ecor}{\end{cor}}   \newcommand{\erem}{\end{rem}}
\newcommand{\econ}{\end{con}}

\renewcommand{\eqref}[1]{(\ref{#1})}

\newcommand{\be}[1]{\begin{equation}\label{#1}}
\newcommand{\bea}[1]{\begin{eqnarray}\label{#1}}
\newcommand{\besn}{\begin{equation*}}
\newcommand{\beasn}{\begin{eqnarray*}}

\newcommand{\dis}{\mathop{\rm d}\nolimits}

      \newcommand{\id}{{1 \mskip -5mu {\rm I}}}

         \renewcommand{\d}{\delta}
    
         \newcommand{\h}{\eta}

         \renewcommand{\L}{\Lambda}
        \renewcommand{\O}{\Omega}

\newcommand{\cA}{\mathcal A}    \newcommand{\cB}{\mathcal B}

    \newcommand{\cL}{\mathcal L}

\newcommand{\cS}{\mathcal S}

     \newcommand{\bR}{\mathbb R}

     \newcommand{\bZ}{\mathbb Z}
\newcommand{\ee}{\end{equation}}
\newcommand{\eea}{\end{eqnarray}}
\newcommand{\ba}[1]{\begin{array}{*{#1}{c}}}
\newcommand{\ea}{\end{array}}

\begin{document}
\title{Dynamical arrest, tracer diffusion and Kinetically Constrained Lattice Gases}

\author{Cristina Toninelli\thanks{Laboratoire de Physique
    Th{\'e}orique de l'ENS, 24 rue Lhomond 75231 PArix Cedex, FRANCE}  and  Giulio Biroli\thanks{Service de Physique Th{\'e}orique, CEA/Saclay-Orme des Merisiers,
F-91191 Gif-sur-Yvette Cedex, FRANCE}}

\maketitle
\begin{abstract}
We analyze the tagged particle diffusion for kinetically constrained
models for glassy systems. We present a method, focusing on the
Kob-Andersen model as an example, which allows to prove lower and
upper bounds for the self diffusion coefficient $D_S$.  
This method leads to the exact density dependence of $D_{S}$, at high density,
for models with finite defects and to prove diffusivity, $D_{S}>0$, at
any finite density for highly cooperative models.
A more general outcome is that under very general assumptions one can exclude
that a dynamical transition, like the one predicted by the
Mode-Coupling-Theory of glasses, takes place at a finite temperature/chemical
potential for systems of interacting particle on a lattice. 
\end{abstract}

\section{Introduction}\label{intro}
Many physical systems, in particular glass forming liquids, display a
very slow dynamics at low temperature/high density \cite{DEBenedetti}. The laboratory glass
transition corresponds to the temperature/density at which the structural
relaxation timescale becomes larger than the experimental one (e.g. one
hour). At this point the glass-forming liquid falls out of equilibrium
and becomes an amorphous rigid material called glass. Thus, the laboratory
glass transition is nothing else than a dynamical crossover and not a
true dynamical transition. However, a natural question is whether
this dramatic increasing of the structural relaxation time is due
to an underlying dynamical transition that takes place at lower
temperature/higher density (but that is unreachable experimentally).
Indeed different analytical approaches, in particular the
Mode-Coupling-Theory (MCT) of the glass transition \cite{MCT} predict
a dynamical arrest at a finite temperature and chemical potential
at which the structural relaxation timescale diverges and the diffusion 
coefficient of a tagged particle, called self-diffusion coefficient $D_{S}$,
vanishes. The interpretation of this transition is based on the 
cage effect: 
particles are trapped in the cage formed by their neighbors 
(however see also \cite{BiroliBouchaud}).  
For particles interacting with a smooth potential 
it is widely accepted that this MCT transition describes at most
a dynamical crossover. Instead in the case of potentials with a  
hard core part, in particular hard sphere systems, there is no agreement.
Experiments on colloids \cite{Pusey}, that can be indeed modeled by hard sphere systems, 
are very often interpreted as if a real dynamical transition took place.
However, in these systems the microscopic timescale is much larger than
for the other glass forming liquids (approximatively nine order of
magnitude), thus it could be argued that one is just looking to the
very first part of the increasing of the relaxation timescale, here
MCT indeed applies but the point at which the dynamical MCT crossover  
takes place is shifted toward unreachable experimental timescales.

What can one say theoretically about the existence of this type of
transition called sometimes dynamical arrest? First let us focus on
the case of interacting particle systems on a lattice. If the
potential between the particle 
is short-range, the Hamiltonian is not singular and the only constraint is the hard core
one, i.e. maximum one particle per site, then it has been proved by Spohn
\cite{Spohn,Sart} that on long time and length scales the tagged particle 
performs a
simple Brownian motion with a self-diffusion coefficient that is positive at any finite
temperature and any finite chemical potential (these models are called in the
mathematical literature symmetric exclusion processes with speed
change and spin exchange dynamics; in the following we'll refer simply
to them as RLG, i.e. reversible lattice gases). Actually, one needs also
the system to be ergodic but physically this is just due to the fact that if there
is a very large correlation length and/or a very large correlation
time than, in principle, the tagged particle will enter in the
Brownian motion regime on larger length and time scales. So at a
critical point the tagged particle could take an infinite time to enter
in the Brownian motion regime. However the lower bound on the self
diffusion coefficient proved in \cite{Spohn,Sart} is valid regardless
of the existence of a critical point. Thus, even if a phase transition
would take place at a temperature $T_{c}$ and a chemical potential
$\mu_{c}$ one gets $\lim_{T\rightarrow T_{c};\mu \rightarrow
\mu_{c}}{D_{S} (T,\mu )}>0$. 

However, the lattice models that are known to display a whole
phenomenology analogous to glass forming liquids and for which 
the existence of a dynamical arrest at finite temperature/chemical potential
has been suggested, are the ones in which further hard constraints besides
 hard core, are imposed. Among the most studied ones there are
the kinetically constrained lattice gases (KCLG) for which  
the jump rates of particles are different from zero not
only if the constraint of having maximum one particle per site is verified 
but {\it also} if some additional constraint is verified, hence the
name kinetically constrained (see \cite{ReviewKLG} for a recent
review). We will comment on the extension of our
work to statically constrained models, like the Lattice Glass Models
introduced in \cite{BiroliMezard}, and more general cases in the conclusion.

Note that the presence of additional hard constraints can change the
physical mechanism behind tagged particle diffusion quite a lot.
As explained in \cite{ReviewKLG} the KCLG are characterized at high density
by the existence of ``defects'' that are the analog of
vacancies for the RLG. 
A rather good understanding has been reached for
models in which defects, that can freely move in an otherwise
completely filled lattice, consist
in a finite number, independent of the particle density, of vacancies. 
Instead, the cases in which the motion of ``defects''  involve a number of vacancies that
diverges when the particle density approaches one, were much less understood.
 Indeed in the literature one can find numerical
simulations suggesting a dynamical transition at a density less than
one \cite{ReviewKLG}. Recently, in collaboration with D.S. Fisher \cite{KAprl}, we have analyzed
one of the most studied model of this type, the Kob-Andersen model
\cite{KA} proving that no dynamical 
transition takes place whatsoever and unveiling what is the mechanism 
inducing the slow dynamics. Our results and techniques apply also to the other
highly cooperative KCLG \cite{KAlong}. 

In this paper we show how one can, under very general assumptions,
generalize the proof of diffusivity, $D_{S}>0$ of \cite{Spohn,Sart} to KCLG (this has been already
done for simple KCLG with finite size defects in \cite{BT}). 
Our aim is twofold. First, we want to show that under very
general assumptions (see next sections and conclusion) a dynamical
arrest cannot take place at finite temperature/chemical potential even
if there are hard constraints other than the hard core one. 
Second, we want to present a rigorous technique and a method that allows 
to obtain upper and lower bounds on the self-diffusion coefficient 
for KCLG (or more general interacting particle systems).
The usual techniques applied to KCLG in order to obtain predictions
on $D_{S}$, as diagrammatic resummation or approximate
closure of exact equations (see \cite{ReviewKLG,PittsAndersen}), are
completely out of control in the high density regime. 
Although some of them work well in the intermediate density regime 
compared to the results of numerical simulations, they fail in general
at high density and, often, predict a spurious dynamical transition.
Thus, in this context it is particularly important to have 
rigorous methods that allow one to obtain solid analytical
predictions. We will focus on the Kob-Andersen model as an example
but our method can be applied to also other KCLG or more general
interacting particle systems.

Let us finally comment on the continuum case. We are not aware of any
proof of diffusivity at low temperature/high density in the case of Hamiltonian dynamics.
Instead in the case of interacting Brownian particles, even with hard
core (a good model for colloidal systems), it
has been recently shown that the self diffusion coefficient is
positive at any finite temperature/chemical potential
under very general assumptions \cite{BrownianDS}. 
The proof works for hard spheres, Lennard-Jones potential, etc\dots
and it is, as our proof, a generalization of the Spohn's proof
\cite{Sart,Spohn} for RLG. 

{\it Organization of the paper}\\
We think that the technique we will make use of is not
well known in the community working on KCLG therefore we have written this
article in a detailed and self-contained way. The expert reader 
may skip section \ref{taggedparticle} and quickly go through 
section  \ref{triangular}.\\
We introduce in detail the KCLG in section \ref{kcm} and we explain in
detail the KA model in section \ref{KAdefinition}.
In section \ref{taggedparticle} we recall some probabilistic
techniques that have been used to prove bounds on the self diffusion 
coefficient for the RLG \cite{Sart}. In section  
\ref{triangular}, focusing on the KA model on a triangular lattice as
an example, we explain how one can obtain the high density behavior of
the self diffusion coefficient for KCLG with ``defects'' formed 
by a finite number of vacancies.
This section is useful to introduce the notation and the method
before facing the difficult case of highly cooperative KCLG. 
In section \ref{square}, focusing on the KA model on a square lattice, 
we explain how one can obtain strictly positive lower
bounds in the case of highly cooperative KCLG. 
Finally, in \ref{conclusion} we present a final discussion of our results.

\section{\bf Kinetically constrained lattice gases}

\label{kcm}

In the last twenty years there has been a growing interest in
kinetically constrained lattice gases. These were introduced as models for
supercooled liquids close to the glass transition
\cite{FA} and nowadays they are also studied as paradigm 
for general glassy systems \cite{ReviewKLG}.
KCLG are (apparently) similar to RLG. As discussed before
there is however an important difference in the choice of the jump
rates of particles that are different from zero not
only if the constraint of having maximum one particle per site is verified 
but {\it also} if some additional constraint is verified, hence the
name kinetically constrained. 
This choice of jump rates was originally devised in order to mimic the cage
effect, that might be at the heart of the glassy behavior and the slow
dynamics of glass forming liquids. Indeed a 
molecule in a dense liquid is typically trapped in a cage created
by surrounding particles (see \cite{Weeks} for a visual experimental example) and
this takes place in the regime of temperature and density at which
the dynamics slows down dramatically. 

More specifically, kinetically constrained lattice gases are stochastic lattice gases
 with hard core exclusion, i.e. systems of particles on a lattice
 $\Lambda$ with the constraint that on each site there can be at most
 one particle.  A configuration is therefore defined by giving for
 each site $x\in\Lambda$ the occupation number $\eta_x\in \{0,1\}$,
 which represents an empty or occupied site respectively.
 The dynamics is given by
 a continuous time Markov process on the configuration
 space $\O_\Lambda=\{0,1\}^{|\Lambda|}$ which consists of a sequence of particle
 jumps.
 A particle at site $x$ attempts to jump to a different site $y$
 with a fixed rate $c_{x,y}(\eta)$, which in general
  depends both on $\{x,y\}$ and on the
 configuration $\eta$ over the entire lattice. The discretized time version
 of the process is the following. At time $t$ choose at random a particle, let $x$ be its position,
 and a site $y$. At time $t+dt$, the particle has jumped from $x$ to $y$ with probability $c_{xy}(\eta(t))$,
 while with probability $1-c_{xy}(\eta(t))$ the configuration has remained unchanged.
In other words, the probability measure at time t, $\mu_t$, can be derived
by the initial measure $\mu_0$ as

\begin{equation}
\mu_t(\eta)=\sum_{\eta'\in\{0,1\}^{|\L|}}\exp\left(\cL t\right)\mu_0(\eta')
\label{evolution}
\end{equation}
where  $\cL$, the {\sl generator of the Markov process}, is the operator
which acts on local functions
 $f:\O_\L \to \bR$ as

\begin{equation}
\label{generator}
\cL f \, (\eta)= \sum_{\{x,y\}\subset\Lambda }
c_{x,y}(\eta) \left(f (\eta^{xy})-f(\eta)\right)
\end{equation}
where we defined

\begin{equation}
\label{exchange}
(\eta^{xy})_z :=
\left\{
\begin{array}{ll}
\eta_y & \textrm{ if \ } z=x\\
\eta_x & \textrm{ if \ } z=y\\
\eta_z  & \textrm{ if \ } z\neq x,y
\end{array}
\right.
\end{equation}
The simplest model is the simple symmetric exclusion process, SSEP, in
which $c_{x,y}^{SSEP}(\eta)=\eta_x(1-\eta_y)+\eta_y(1-\eta_x)$ for nearest
neighbors $\{x,y\}$, $c_{x,y}(\eta)=0$ otherwise.  Therefore, only nearest
neighbor jumps are allowed and there are no further kinetical constraints
besides hard core. The definition {\sl kinetically constrained} refers
more properly to models in which jump rates impose additional
requirements in order for the nearest neighbor move to be allowed. In other
words, the rate $c_{x,y}(\eta)$ can be zero
 for some choices of the
configuration $\eta$ and the couple $\{x,y\}$ even
 if $\eta_x=1$ $\eta_y=0$, thus preventing the jump of a particle from
site $x$ to final empty site $y$.
From the above definition it is immediate to see that dynamics preserves
the number of particles, i.e. the hyperplanes with fixed number N of
particles  $\O_{\L,N}:=\{\h\in\O_\L\,\:  \sum_{x\in\L}\h_x=N\}$ are
invariant under dynamics. Moreover,  in general 
the rates are chosen in order to
satisfy detailed balance w.r.t. the uniform measure $\nu_ {\L,N}$
 on such hyperplanes.
  In other
 words condition
 $c_{x,y}(\eta)=c_{y,x}(\eta^{xy})$ is satisfied for any choice of $\eta$ and the couple $\{x,y\}$.
This implies  that the generator is reversible with respect to $\nu_{\L,N}$
 and therefore
 $\nu_{\L,N}$ is stationary\footnote{Let $\mu(g,h)=\sum_{\eta\in\O}\mu(\eta)g(\eta)h(\eta)$.
 $\cL$ is reversible with respect to $\mu$ if, for any functions $f$ and $g$, equality $\mu(g,\cL f)=\mu(f,\cL g)$
 holds. By a direct calculation it is possible to check that
  detailed balance implies reversibility with respect to $\nu_{\L,N}$, therefore
  the choice $g(\eta)=1$ $\forall \eta$, implies
 $\mu(\cL f)=0$ $\forall f$. This, together with (\ref{evolution}) implies that
 $\nu_{\L,N}$ is invariant under time evolution.}.  Note that $\nu_{\L,N}$ is nothing else than
  canonical measure with zero Hamiltonian, i.e. with this choice of the rules there
  are no static interactions beyond hard core and an equilibrium transition cannot occur.
  However, a priori it is possible that a dynamical ergodic/non--ergodic 
transition occurs
 for some choices of the rules. 
To our knowledge, the KCLG which have been
considered so far do not display such transition (see \cite{KAprl,KAlong} 
for the proof that an ergodic/non--ergodic transition does not occur 
in some highly cooperative KCLG). On the other hand, such
transition has  
been proved to
occur  for a kinetically constrained spin model, 
namely North-East model \cite{NorthEast, NE2}.
  We shall discuss in the following possible forms of such transition
rates.
  However, we emphasize since now that
  the degeneracy of
  the rates implies that
   $\nu_{\L,N}$
  is not the unique invariant measure, i.e. the system is not ergodic on
$\O_{\L,N}$ and this will have several consequences on dynamics inducing a
very different behavior with respect to RLG case.

\section{Definition of the Kob-Andersen model}
\label{KAdefinition}

 The KA model is a kinetically constrained lattice model with jump rates

\begin{equation}
\label{KA2}
c_{x,y} (\eta) :=
\left\{
\begin{array}{ll}
c_{x,y}^{SSEP}(\eta) & \textrm{ if \ }  \sum_{ \genfrac{}{}{0pt}{1}{z\in\Lambda,z\neq y }{\dis(x,z)=1}}\eta_z\leq m \textrm{\ and \ } \sum_{ \genfrac{}{}{0pt}{1}{z\in\Lambda,z\neq x }{\dis(y,z)=1}}\eta_z\leq m
\\
0 & \textrm{ otherwise \ }
\end{array}
\right.
\end{equation}
namely a particle can move only if the hard core constraint is verified,
as for the SSEP, {\it and}
only if both before and after the move it
 has no more than $m$ neighboring particles.
If $\Lambda$ is an hypercubic $d-$dimensional lattice $m$ will take
values only from $0$ to $2d-1$ (different values of $m$ define
different KA models).
 Note that for $m=2d-1$ the simple symmetric
 exclusion case with Hamiltonian equal to zero is recovered.
 For future purposes it is useful to reformulate the rule
 in term of motion of vacancies. Indeed, as can be  easily  verified, the above definition corresponds to
 {\it vacancies} moving only if the
 initial and final sites have at least $s=z-m-1$ neighboring
 vacancies, with $z=2d$ the coordination number of the lattice.
Therefore the model is completely defined by the choice of the couple $d,m$ or equivalently $d,s$.

Note that these rates satisfy detailed balance with respect to $\nu_{\L,N}$, i.e. uniform measure
on the hyperplanes with fixed number of particles. However there exist
configurations that are blocked under the dynamics, $\nu_{\L,N}$ is not the unique invariant measure on the hyperplane
and the process is never ergodic. 
For example, in the case $d=2$ $s=1$ with periodic boundary condition, a configuration which has
a double row of sites completely filled belongs to a different ergodic component with respect to
any other configuration which does not contain such structure. 
Indeed, one can directly check that the particles
belonging to the double row can never move.

On the infinite lattice $\L=\bZ^d$ the model satisfies detailed
balance with respect to the Bernoulli
product measure $\mu_{\bZ^d, \rho}$ at any density $\rho$. It is
immediate to check that for
$s\geq d$ the system is not ergodic at any density $\rho>0$. Indeed in
this case
all d--dimensional hypercubes of any size which are
completely occupied by particles are blocked forever. On the other
hand,
in \cite{KAprl,KAlong}, we have proved that for $s<d$ the model is
ergodic at any density $\rho\in[0,1]$ with probability one.
As a by product of the ergodicity proof we have established the following property, which
will be a key ingredient for the proof in section \ref{square}.
Let $p$ be a positive number such that $p<1$.
For any fixed density $\rho<1$ 
there exists a finite length $\Xi(\rho)$
such that, by sorting at random with probability $\mu_{\L,\rho}$ 
a configurations on a hypercubic d--dimensional lattice $\L$ of linear size $\Xi(\rho)$,
with probability greater than $p$ this configuration is such that any
particle exchange inside a finite box around the origin can be performed through a
suitable path of allowed moves\footnote{
 The length $\Xi(\rho)$ depends on the choice of $d$, $s$ and the size
of the box, we drop the dependence on these parameters for simplicity
of notation.}. Furthermore $p$ can be taken arbitrary close to one
taking a suitable (large but finite) $\Xi(\rho)$.
In the following we let a
configuration with such property be a {\sl frameable configuration} and
refer to \cite{KAlong,KAprl} for the demonstration of above property and an
explanation of the chosen name.

\section{\bf Diffusion of the tagged particle}
\label{taggedparticle}

Consider a kinetically constrained model on the infinite lattice  $\Lambda=\bZ^d$
and start at time zero from the equilibrium distribution, so that the process will be stationary.
Then single out one particle, the tracer, and analyze its motion.
In the density regime where the process is ergodic one can repeat the arguments in \cite{KV,Sart} and show that
under a diffusive rescaling the position of the tracer at time $t$,  $\vec{x}(t)$, converges to a Brownian motion
with self diffusion matrix $D_S(\rho)$. More precisely,
 $\lim_{ \epsilon\to 0} \epsilon \vec {x}(\epsilon^{-2}t)=\sqrt{2D_S}~\vec{b}(t)$, where $\vec{b}(t)$ is standard
 Brownian motion 
and
  the self diffusion matrix $D_S(\rho)$ is given by the variational
  formula \cite{Spohn}\footnote{Note that by the identity
(\ref{variational}) one obtain that the 
matrix $D_S$ is positive,
therefore the square root is a well defined function.}:
\begin{eqnarray}
&& (\vec{l}{\cdot} D_S(\rho){\cdot}\vec{l})
=\inf_{f}\!\! \left[\frac{1}{2}\!\sum_{\genfrac{}{}{0pt}{1}{\{y\neq 0\}\subset\Lambda }{\phantom{x\neq 0,y\neq 0}}}\!\!
\mu_{\rho,0}(c_{0,y}(\eta)(1-\eta(y))[\sum_{i=1}^{d}(\vec{l}{\cdot}\vec{d})+\right.\nonumber\\
&& \left. \left.f(\tau_{-y}\eta^{0y})-f(\eta)]^2\right) 
+\!\frac{1}{4}
\sum_{\genfrac{}{}{0pt}{1}{\{x,y\}\subset\Lambda }{x\neq 0,y\neq 0}} \mu_{\rho,0}\left(c_{x,y}(\eta)
[f(\eta^{xy})-f(\eta)]^2\right)
\right]\nonumber\\
&&
\label{variational}
\end{eqnarray}
where $\vec {l}$ is a unit vector in $\bZ^d$,
$(\vec{a}{\cdot}\vec{b})$ is Euclidean scalar product, $\tau_{-y}\eta$ is
the configuration obtained translating the configuration $\eta$  of $y$,
$\mu_{\rho,0}$ is the
  Bernoulli measure at density $\rho$ conditioned to the existence  of a particle in the origin and
 the infimum is over
all real--valued local functions $f$, i.e. functions which depend on a
finite number of occupation variables.
For spatially isotropic systems, as the one we will deal with, the self
diffusion matrix is usually proportional to identity, the proportionality coefficient
being the so--called {\sl self diffusion coefficient}. For future purposes it is useful to
define the two sums in (\ref{variational}) as $D_1(\rho,f)$ and
$D_2(\rho,f)$

\begin{eqnarray}
D_1(\rho,f)&=&\frac{1}{2}\sum_{\{y\neq 0\}\subset\Lambda }
 \mu_{\rho,0}(c_{0,y}(\eta)(1-\eta
(y))[\sum_{i=1}^{d}(\vec{l}{\cdot}\vec{d})+\nonumber\\
&&\left.f(\tau_{-y}\eta^{0y})-f(\eta)]^2\right)\label{d1}\\
 D_2(\rho,f)&=&\frac{1}{4}
\sum_{\genfrac{}{}{0pt}{1}{\{x,y\}\subset\Lambda }{x\neq 0,y\neq 0}} \mu_{\rho,0}\left(
c_{x,y}(\eta)[f(\eta^{xy})-f(\eta)]^2\right)
\label {d2}
\end{eqnarray}
It is immediate to notice that $D_1(\rho,f)\geq0$ and $D_2(\rho,f)\geq 0$
at any density $\rho$, and therefore $D_S(\rho)\geq 0$. However, for the
tagged particle process to be diffusive, $D_S$ should be strictly positive.
For the RLG case in any dimension $d\geq 2$ the  result  $D_S>0$ 
at any density $\rho<1$ holds.
Since the proof  \cite{Sart} 
uses as a key ingredient the fact that jumps from occupied to neighboring
empty sites are always allowed, it  cannot be trivially extended to
kinetically constrained models. We emphasize that this is not only a
technical difficulty. Indeed, due to the presence of kinetic constraints, the
physical mechanism behind the diffusion of the tagged particle may be
rather different as we shall show in the next sections.

In the following section we will prove that $D_S>0$ at any $\rho<1$
for KA model on a triangular lattice. 
For this choice of the lattice there exist ``defects'', namely clusters of
two vacancies that can freely move through an
otherwise totally filled lattice and such that the tagged particle can move
whenever such ``defects'' passes by. 
Therefore these defects play the
same role of the vacancies for the RLG case and 
 the above mentioned proof of diffusivity 
 (\cite{Sart}, see also \cite{Spohn})) can be generalized.
In the following we sketch this result in some detail since 
it can be extended to other models with finite ``defects'' and it is useful 
to introduce the notation and the method before analyzing the more difficult case
of KA model on hypercubic lattices. Indeed, in this case it is not
possible to construct 
 finite cluster of vacancies that can freely move into an
otherwise totally filled lattice and additional work is required
to prove diffusivity.
As explained in \cite{KAprl}, in this case the mechanism behind
diffusion is based on the existence of
``quasi--defects'', namely cluster of vacancies 
which can move into typical
regions of the system and allow the motion of a tagged particle (via a
particular path of elementary moves) when they pass by. 
The size of such defects is density dependent
and diverges for $\rho\to 1$, which makes the motion {\sl highly cooperative}. 

\section {\bf A model with finite size defects: KA $s=1$ on 
a triangular lattice}
\label{triangular}

\begin{figure}[bt]
\centerline{
\includegraphics[width=0.5\columnwidth]{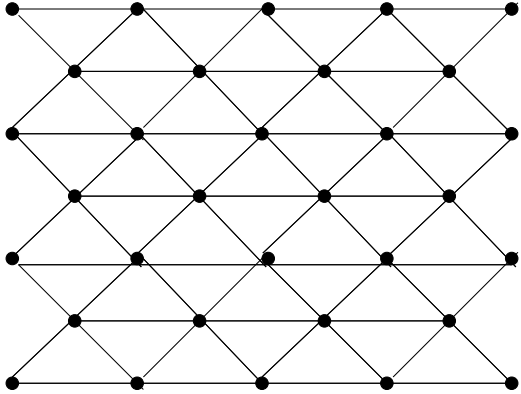}}
\caption{ The triangular lattice.}
\label{tria1}
\end{figure}

\begin{figure}[bt]
\centerline{
\includegraphics[width=0.9\columnwidth]{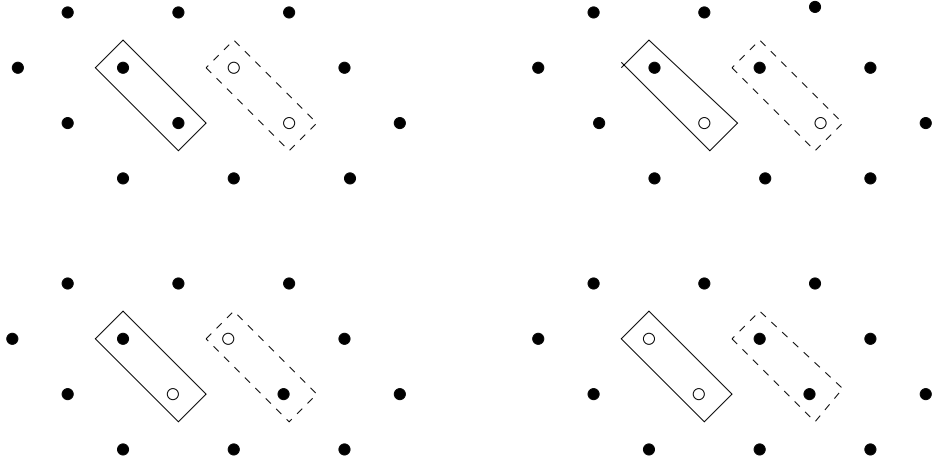}}
\caption{Sequence of moves which allow the displacement of a couple of
  neighboring vacancies. Circles denotes empty sites, filled dots
  stand for occupied sites}
\label{basic}
\end{figure}

Consider the KA model with $s=1$ on a triangular 
lattice $\Lambda$ represented in figure \ref{tria1}. 
More precisely, the set of sites in $\Lambda$ is
the union of sites in a square  lattice $\Lambda_1$ and in its dual
$\Lambda_2$, i.e. the lattice obtained by displacing $\Lambda_1$ of
$(e_1+e_2)/2$ with $e_1=(1,0)$, $e_2=(0,1)$. Furthermore,
two sites $\{x,y\}\in\Lambda$ are nearest
neighbors if $x-y=\pm e_1$ or $x-y=\pm (e_1+e_2)/2$ or
else  $x-y=\pm (-e_1+e_2)/2$.
As already noticed, rates (\ref{KA2}) reformulated in terms of
vacancies correspond to the rule that a vacancy can move only if the
initial and final sites have at least $s$ neighboring vacancies, where $s=1$
in the case we are considering. In the triangular lattice 
two neighboring sites always share a common third neighbor.  
Therefore any of two nearest neighbor vacancies can move to the
common third neighbor. In other words a set of two neighboring
vacancies, which we call ``a defect'',
can be freely moved into an otherwise totally filled
lattice, as can be immediately checked (see figure \ref{basic}). 
\subsection{Heuristic arguments and upper bound on the self diffusion coefficient}\label{heuristic}
There is a simple heuristic argument based on the independent motion of these defects
that leads to the correct density dependence of the self-diffusion
coefficient at high density. Call $\rho_{d}$ the density of defects
and $\tau_{d}$ the timescale on which defects move. The self-diffusion
coefficient $D_{s}$ of a tagged particle is expected to be proportional to the
inverse of the time $\tau_{p}$ on which each particle moves of one
step. On the timescale $\tau_{d}$ the number of particles that have
jumped is of the order $V\rho_{d}$ where V is the total number of
sites. Thus we find
\begin{equation}\label{}
\frac{\tau_{p}}{\tau_{d}}V\rho_{d}\propto V\rho.
\end{equation}
As a consequence, at density close to one, we get $D_{S}\propto
\rho_{d}/\tau_{d}$. Note that we have assumed that the size of the
defects do not change, in particular do not diverge in the limit $\rho
\rightarrow 1$, otherwise the reasoning has to be changed slightly.
Since for the KA model on the triangular lattice we are focusing on
the defects are formed by two vacancies we find, 
in the limit $\rho \rightarrow 1$, 
$\rho_{d}\propto (1-\rho
)^{2}$ and $\tau_{d}\propto O (1)$, hence, $D_{S}\propto (1-\rho
)^{2}$ (similarly in the SSEP case a defect is just a vacancy and 
one gets $D_{S}\propto (1-\rho )$).

In the following we will show how, using the existence of these defects,
it is possible to  generalize  the proof of the RLG
case \cite{Sart} and show that indeed $D_{S}\propto (1-\rho)^{2}$. In principle
our procedure is generalizable to all cases with defects having a size
which doesn't diverge in the limit $\rho \rightarrow 1$.
To obtain $D_{S}\propto (1-\rho)^{2}$ we shall prove that when $\rho $
is close enough to one $D_{S}$
is bounded from above and below respectively by $K_{U}(1-\rho)^{2}$ 
and $K_{L}(1-\rho)^{2}$ where $K_{U,L}$ are two positive constants 
(of course $K_{U}\geq
K_{L}$). 

The proof of the upper bound is very easy: it consists just in
choosing an appropriate test function $f (\eta)$ and evaluating 
the term in the parenthesis in the variational formula (\ref{variational}).
Indeed consider the test function $f_{0}(\eta)=\eta_0$, which for each
configuration is equal to the occupation number in the origin. 
Plugging it into (\ref{variational}) we find a term which is proportional
to the probability to have a vacancy on a fixed site and at least another
vacancy close to it (this is nothing else that the probability to have
a defect). Thus we obtain $D_{S}\le K_{U} (1-\rho
)^{2}$ where $K_{U}$ is a suitable positive constant.  
\subsection{Lower bound on the self-diffusion coefficient}\label{lowertriangular}
Now let us focus on the lower bound which is more involved. 
Instead of dealing directly
with the variational formula (\ref{variational}), we define a proper
auxiliary model and proceed in two steps. First, we establish that
$D^{aux}_S>0$; second, we prove that $D_S\geq c D^{aux}_S$ with $c$ a
positive constant. More precisely, we will introduce an auxiliary
process and prove that the diffusion coefficient in direction $e_1$ is
positive, i.e. $e_1 D_s^{aux} e_1>0$ and $e_1 D_S e_1\geq e_1
D_S^{aux} e_1$. In an analogous way one can then introduce an
auxiliary process to show the same inequalities in direction
$(e_2+e_1)$, which completes the proof of the positivity of the self
diffusion matrix.
\subsubsection{Construction of the auxiliary process}\label{auxtriang}
\begin{figure}[bt]
\psfrag{x}[][]{$x$}
\psfrag{R1}[][]{$R_x^1$}
\psfrag{R2}[][]{$R_x^2$}
\centerline{
\includegraphics[width=0.5\columnwidth]{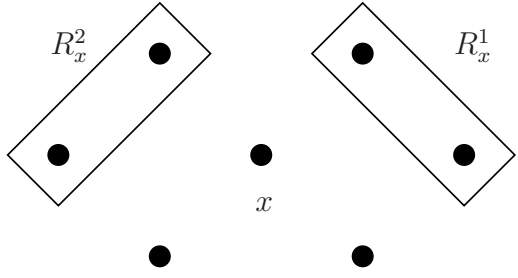}}
\caption{Sites inside the closed line correspond, form left to right,
to sets $R_x^{1}$ and $R_x^2$.}
\label{R}
\end{figure}

Let us introduce some notation.
Consider
 the following subsets of $\L$

\begin{eqnarray}
R_x^{1}
&:=&
\big\{x+ e_1, ~~x+\frac{e_1+e_2}{2} \big\}\nonumber\\
R_x^{ 2}
&:=&
\big\{(x- e_1, ~~x+\frac{-e_1+e_2}{2}\big\}
\end{eqnarray}
namely $R_x^i$ are the two couples of neighboring sites of $x$
represented in figure \ref{R}.

We next define $\eta^{1,2,x}$
as the configuration obtained from $\eta$ by exchanging the occupation numbers
in $R^{1}_{x}$ with the corresponding ones in $R^{2}_{x}$, namely

\begin{equation}
\label{exch}
\big(\eta^{1,2,x}\big)_z :=
\left\{
\begin{array}{ll}
\eta_{x-e_1} & \textrm{ if \ } z=x+ e_1\\
\eta_{x+e_1} & \textrm{ if \ } z=x- e_1\\
\eta_{x+ \frac{e_1+e_2}{2}} & \textrm{ if \ } z=x+ \frac{e_1+e_2}{2}\\
\eta_{x+ \frac{-e_1+e_2}{2}} & \textrm{ if \ } z=x+ \frac{-e_1+e_2}{2}\\
\eta_z & \textrm{ if \ } z\not\in R_x^{1}\cup R_x^{2}
\end{array}
\right.
\end{equation}
Finally, we introduce the events

\begin{equation}
\cA_x^i:= \big\{\eta \in\O \,:\: \eta_z=0\: \forall z\in R_x^i \big\}
~~~~~~~ \cA_x:=\cA_x^1\cup\cA_x^2
\end{equation}
which contain
 configurations having the two neighboring sites in $R_x^i$
empty, namely configuration with a defect in $R_x^i$.

Let us denote by $x$ the position of the tagged particle and
define the auxiliary process as follows.
At time $t=0$ the tagged particle is at the origin and there is a
``defect'' in a neighboring couple of sites, namely $x(0)=0$
and $\eta(0)\in\cA_x$.
(i.e. $\eta_{e_1}(0)=\eta_{e_2/2+e_1/2}(0)=0$
or $\eta_{-e_1}(0)=\eta_{-e_1/2+e_2/2}(0)=0$). 
Then the process evolves as follows
\begin{description}
\item (i) The tagged particles can jump from $x$  to
 $x+e_1$. The rate of the jump is one  
if $\eta\in\cA_x^1$ (i.e. $\eta_{x+e_1}=0$  and $\eta_{x-e_1/2+e_2/2}=0$);
\item (ii) The tagged particles can jump from $x$  to  $x-e_1$. 
The rate of the jump is one  
if $\eta\in\cA_x^2$ (i.e. $\eta_{x-e_1}=0$ and $\eta_{x-e_1/2+e_2/2}=0$);
\item (iii) Configuration $\eta$ is exchanged to $\eta^{1,2,x}$
(i.e. the occupation variables of sites $x+e_1$, $x+e_1/2+e_2/2$
 are exchanged with
  the occupation variable in $x-e_1$ and $x-e_1/2+e_2/2$, respectively).
 The rate of the exchange is one;
\item(iv) All other moves are not allowed.
\end{description}
With this definition of the rules
it is immediate to check that, since the configuration 
at time zero is such that $\eta(0)\in\cA_{x(0)}$, 
at any subsequent time $t>0$ condition
$\eta(t)\in\cA_{x(t)}$
will hold. Moreover, since the defect can be moved from $R_x^1$ to $R_x^2$ 
and viceversa (move (iii)
above), the jump of the
tagged particle to any of the two neighbors in direction $e_1$ can always
be performed via a sequence of at most two moves. 
Therefore the self diffusion coefficient in direction
$e_1$ is strictly greater than zero  at any density $\rho<1$, namely
$e_1 D_S^{aux} e_1>0$. This can be rigorously proved in the same way as for RLG
(see \cite{Spohn}). 
\subsubsection{Proof of the inequality between $D_{S}$ and $D_{S}^{aux}$}\label{auxtriang2}
Let us now turn to the second step, namely establishing inequality
$D_S>cD_S^{aux}$, with $c>0$. Since move (iii) for the auxiliary process is
not allowed for KA, some work is required to establish such inequality.
The basic idea is to show that all the moves allowed for the auxiliary
process can be performed through a proper finite sequence of elementary
nearest neighbors jumps which are  allowed for the original model. If this is
true then it is natural to expect that the above inequality among the diffusion
coefficients can be rigorously established. Let us outline the
proof. Consider the second term, $D_2$, of the variational formula
(\ref{variational}). The above choice of the rates yields for the auxiliary process

\begin{equation}
\label{variational2aux}
D_2^{aux}(f,\rho)=\frac{1}{4\mu_{\rho,0}(\cA_x)}<\id_{\cA_{x}}(\eta)\left(f(\eta^{1,2,x})-f(\eta)\right)^2>_0
\end{equation}
As already mentioned any couple of neighboring vacancies can be freely
moved through the lattice using elementary moves allowed by KA, see
figure \ref{basic}. 
In particular, thanks to the fact that the $\eta$ which enters
in (\ref{variational2aux}) have a couple of vacancies either in the set
$R_x^1$ or $R_x^2$ (or both), by using the basic moves in figure \ref{basic} 
it is possible to construct a 
 path of elementary nearest neighbor exchanges which have unit rate for KA
model and connect $\eta$ to $\eta^{1,2,x}$. Moreover, such path is independent
on the configuration outside $R_x^1$ and $R_x^2$.
In other words, by recalling
definition (\ref{exchange}) and letting the exchange operator $T_{x,y}$ be

\begin{equation}
T_{x,y}\eta=\eta^{x,y}
\end{equation}
the following equivalence holds

\begin{equation}
\label{minimalpath}
\eta^{1,2,x}=\prod_{j=1}^n T_{x_{j+1},x_j}\eta
\end{equation}
where $n$ is the length of the above defined path and
$\eta\to\eta^{x_{j+1},x_j}$
is the elementary exchange which constitute the j--th move of the
path.
Therefore, term $f(\eta^{1,2,x})-f(\eta)$ in
(\ref{variational2aux}) can be rewritten by a telescopic sum as

\begin{eqnarray}
\label{telescopic}
f(\eta^{1,2,x})-f(\eta)&=&\sum_{i=1}^n\left(f(T_{x_{i+1},x_i}\eta_i)-f(\eta_i)\right)
\end{eqnarray}
where we have defined $\eta_1,\dots \eta_n$ as $\eta_1\equiv\eta$,
$\eta_i\equiv\eta_{i-1}^{x_i,x_{i-1}}$.
Then, by
using (\ref{telescopic}) and Cauchy--Scharwz inequality we obtain

\begin{equation}
\label{telescopic2}
\left( f(\eta^{1,2,x})-f(\eta)\right)^2 \leq n \sum_{j=1}^n
\left(f(T_{x_{j+1},x_{j}}\eta_j)-f(\eta_j)\right)^2
\end{equation}
Since the path $\eta_1,\dots \eta_n$ has been chosen in order that the
exchanges in the right hand side are all elementary exchanges
allowed for KA process, i.e. with unit rate,
we can rewrite the above inequality by
introducing in the right hand side the corresponding jump rates 

\begin{equation}
\label{telescopic3}
\left( f(\eta^{1,2,x})-f(\eta)\right)^2 \leq n \sum_{j=1}^n 
c_{x_{j+1},x_{j}}(\eta_j)
\left(f(T_{x_j,x_{j+1}}\eta_j)-f(\eta_j)\right)^2
\end{equation}
Inserting the above inequality in (\ref{variational2aux}), using the
change of variables $\eta_i\to\eta$ and the invariance of equilibrium
measure under exchange of occupation numbers
yields, for any
real--valued local function $f$,

\begin{eqnarray}
\label{eq2}
D_2^{aux}(f,\rho)&=&\frac{1}{4\mu_{\rho,0}(\cA^x)}<\id_{\cA_{x}}(\eta)\left(f(\eta^{1,2,x})-f(\eta)\right)^2>_0\nonumber\\
 &\leq & \frac{1}{4\mu_{\rho,0}(\cA^x)} ~n~ {\cal{N}} \sum_{ \genfrac{}{}{0pt}{1}{\{x,y\}\subset\Lambda
  }{\dis(x,y)=1} } <c_{x,y}(\eta)\left(f(\eta^{x,y})-f(\eta)\right)^2>_0\nonumber\\
&&
\end{eqnarray}
where we let ${\cal{N}}$ be the maximal number of times a single exchange has
been used in the path. In other words, let ${\cal{N}}_{ij}$ be the
number of times operator $T^{ij}$ appears in expression \ref{minimalpath}, we let
${\cal{N}}:={\mbox{max}}_{\{ij\}}{\cal{N}}_{ij}$ where the maximum is
taken over all the couples of neighboring $\{i,j\}$. We emphasize that
${\cal{N}}$ and $n$ are independent on the choice of the
configuration, since the path is
 fixed once for all and does not depend on the value of the
configuration outside $\cA_x$.
>From the above inequality and recalling definition
 (\ref{variational}) it is immediate to conclude that

\begin{equation}
D_2(\rho,f)\geq \frac{\mu_{\rho,0}(\cA^x)}{n{\cal{N}}} D_2^{aux}(\rho,f)
\end{equation}
for any $\rho$ and $f$. 
On the other hand,
since the rates of the moves for  the tagged particle in the auxiliary process are
always smaller or equal to the correspondent rates for KA (see moves
(i) and (iii)), the following inequality trivially holds among the first term of the variational expression
(\ref{variational}) for $D_S$ for the two processes

\begin{equation}
D_1(\rho,f)\geq \mu_{\rho,0}(\cA^x)D_1^{aux}(\rho,f)
\end{equation}
for any $\rho$ and $f$. Therefore

\begin{equation}
D_S(\rho)\geq\frac {\mu_{\rho,0}(\cA^x)}{n{\cal{N}}}~D^{aux}(\rho)\geq
c\mu_{\rho,0}(\cA^x)~D^{aux}(\rho)
\end{equation}
with $c$ a strictly positive constant and
$\mu_{\rho,0}(\cA^x)=(1-\rho^4-4\rho^3(1-\rho)-4\rho^2(1-\rho)^2)$,
which is also strictly positive
at any density $\rho<1$. In particular, in the high density limit,
$\mu_{\rho,0}(\cA^x)\propto (1-\rho)^2$, hence, we finally get 
$D_{S}\geq K_{L} (1-\rho )^{2}$ with $K_{L}$ a positive constant.

\section{ \bf Self diffusion coefficient for highly cooperative KCLG}
\label{square}
In the following we analyze the $s=1$ KA model on a square
lattice. This is a case in which it is impossible to identify defects
with a density independent size that can freely move inside the
lattice. Thus, the key ingredient for the diffusivity proof discussed
in the previous section does not hold. This is not a simple technical difficulty
but it is deeply related to the fact that diffusion takes place in a
different way in this case. As it has been found in
\cite{KAprl} by analytical and numerical arguments the diffusion takes
place thanks to the cooperative motion of a number
of vacancies that diverges approaching unit density. This leads to 
an extremely rapid decreasing of the self-diffusion coefficient
that, however, remains positive until unit density.\\
In the following we want to show how one can prove that indeed 
$D_{S}$ remains positive for $\rho <1$. Our procedure is very general
and can be applied to the other highly cooperative KA models as well
as to many other interacting lattice particle systems.
Note that in this case, at variance with what done in the previous
section, we will not discuss the upper bound on $D_{S}$. Mainly because 
we do not expect that the lower bound we establish does give the
right density dependence (see \cite{KAprl,KAlong}). 

The strategy of the proof is similar to the one discussed in the
previous section: we introduce a proper auxiliary process such that 
$D_S>cD_S^{aux} $ with $c>0$ and then we prove that $D_{S}^{aux}>0$.
The key physical ingredient that allows us to find that such an
auxiliary process exists is that (\cite{KAprl,KAlong} and section
\ref{KAdefinition}) all the particle exchanges inside a finite box
around the origin can be performed with a very high probability $p$ through a suitable path in
configuration space that involves particles at most at distance $\Xi(\rho)$
from the origin. In this case the restricted configuration in the
sublattice $\L_{\Xi}$ of size $\Xi$ around the origin is called {\sl
  frameable} (see section  \ref{KAdefinition}). 
Furthermore $p$ can be chosen arbitrary close
to one taking a suitably large $\Xi(\rho)$.
If this property is verified then it's possible to find an auxiliary
process that maps onto a random walk in a random environment and has
$D_{S}^{aux}>0$.

Let us explain the idea in more detail.
Consider a configuration on the infinite lattice $\bZ^2$ sorted at random with Bernoulli measure
at density $\rho$ and
focus on a sub-lattice $\L_{\Xi}$ of linear size $\Xi(\rho)$.
As recalled in section (\ref{KAdefinition}), we know that the restriction of the configuration to $\L_{\Xi}$
is frameable with probability  almost one. Therefore in the initial configuration
the tagged particle is with very high probability inside a frameable region of size
$\Xi$. 
Moreover, if one divides the infinite lattice in sub-lattices of linear size $\Xi$,
there exists with unit probability a percolating cluster
of sub-lattices such that the initial configuration restricted to each sub-lattice is frameable.

Thus, roughly speaking, if we define an auxiliary process
such that: (1) the tagged particle can move if it is inside a frameable
square, (2) during the dynamics frameable sublattices remain frameable
and the tagged particle remains always inside the percolating cluster
then we can reconstruct any move of such process through a finite
sequence of moves allowed by KA (indeed any nearest neighbor move in a frameable configuration of linear size $\Xi$
can be performed through a sequence of elementary moves
allowed by KA rules and the length of such path is at most of order
$\Xi^2$)and prove inequality $D_{S}>cD_{S}^{aux}$. Furthermore
since in the auxiliary process the particle moves on a percolating
giant cluster then $D_{S}^{aux}>0$.


\subsection{Construction of the auxiliary diffusive process for the KA
$s=1$ on a square lattice }\label{auxiliary}

\begin{figure}[bt]
\centerline{
\includegraphics[width=0.9\columnwidth]{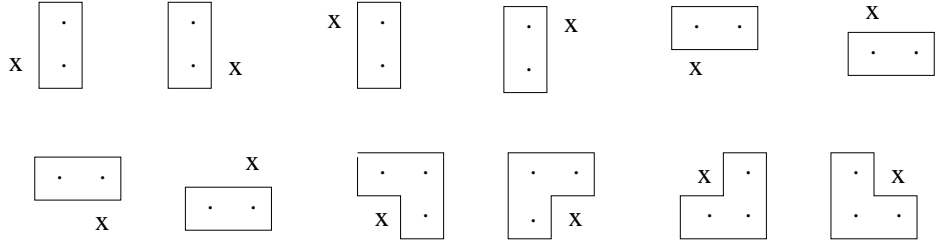}}
\caption{ Sites inside the closed line correspond, form left to right an up down,
to sets $R_x^{+1}, R_x^{-1}, R_x^{+2}, R_x^{-2}, R_x^{+3}, R_x^{-3}, R_x^{+4}, R_x^{-4}, Q_x^{+1},
Q_x^{-1}, Q_x^{+2}, Q_x^{-2}$}
\label{RQ}
\end{figure}

Let us introduce some notation.
Consider
 the following subsets of $\bZ^2$

\begin{eqnarray}
R_x^{({\pm} 1)}
&:=&
\big\{(x_1{\pm} e_1, x_2),(x_1{\pm} e_1, x_2+e_2) \big\}\nonumber\\
R_x^{({\pm} 2)}
&:=&
\big\{(x_1{\pm} e_1, x_2), (x_1{\pm} e_1, x_2-e_2) \big\}\nonumber\\
R_x^{({\pm} 3)}
&:=&
\big\{(x_1, x_2{\pm} e_2), (x_1+e_1, x_2{\pm} e_2) \big\}\nonumber\\
R_x^{({\pm} 4)}
&:=&
\big\{(x_1, x_2{\pm} e_2), (x_1- e_1, x_2{\pm} e_2) \big\}\nonumber\\
Q_x^{({\pm} 1)}
&:=&
\big\{(x_1, x_2+e_2)\cup R_x^{({\pm} 1)} \big\}\nonumber\\
Q_x^{({\pm} 2)}
&:=&
\big\{(x_1, x_2-e_2)\cup  R_x^{({\pm} 2)} \big\}
\end{eqnarray}
$R_x^i$ are the eight possible couples of neighboring sites $\{y,z\}$
such that $y,z\neq x$ and $|y-x|=1$ or $|z-x|=1$ (see figure \ref{RQ}),
in other words one among $y$ and $z$ is neighboring site
to $x$ and the couple does not contain $x$; $Q_x^i$ are the four possible choices of three sites that,
 together with $x$, form a two by two square (see figure \ref{RQ}).
We next define $\eta^{R_x^{+i},R_x^{-i}}$, for $i\in\{ {\pm}1,\dots, {\pm}4
\}$ as the
configuration obtained from $\eta$ by exchanging the occupation numbers
in $R^{+i}_{x}$ with the corresponding ones in $R^{-i}_{x}$
\begin{equation}
\label{exchKA}
\big(\eta^{R_x^{+i}, R_x^{-i}}\big)_z :=
\left\{
\begin{array}{ll}
\eta_{x\mp 2e_1} & \textrm{ if \ } z\in R_x^{{\pm} i}\\
\eta_z & \textrm{ if \ } z\not\in R_x^{+i}\cup R_x^{-i}
\end{array}
\right.
\end{equation}
Then, for $i\in\{{\pm}1,{\pm}2\}$,
we introduce the events

\begin{equation}
\cA_x^i:= \big\{\eta \in\O \,:\: \eta_z=0\: \forall z\in Q_x^i \big\}
\end{equation}
 i.e. configurations having all the sites of set $Q_x^i$ empty
 and
\begin{equation}
\label{cB}
\cB_x^{i} :=
\big\{\eta \in\O \,:\: \eta_{z}=1~~\forall z\in R_x^i,\: \eta\in{\cal{F}}_x^i
\big\}
\end{equation}
for $i\in\{\pm 1,\pm 2,\pm 3, \pm 4\}$, where

\begin{equation}
{\cal{F}}_x^i:
=\big\{\eta \in\O \,:\:\exists \L\in \cS :\: R_x^i\subset\L, ~\d(R_x^i,\partial \L)\geq 3,~\eta|_\L\in {\cal{F}}_\L\big\}
\end{equation}
and $\cS$ is the set of squares in $\bZ^2$ of linear size at most $\Xi$. Here  $\partial\L$ is the boundary of square
sublattice $\L$,
$\eta|_\L$ is the restriction of a configuration to $\L$, ${{\cal{F}}_\L}$ is the set
of frameable configuration in $\L$ and $\delta (A,B)$ is the minimum over the Euclidean distance of all
 the couples $\{x,y\}$ with $x\in A$ and $y\in B$. In other words, $\cB_x^i$ is the set of
 configurations in which a pair of sites adjacent to  $x$ (region $R_x^i$) is filled and 
is internal to a frameable square of linear size at most $\Xi$.
Let $\eta(0)$ be the configuration and $x(0)$ the position of the tagged particle
at time zero, we define $\bar{\eta}(0)$ as

\begin{equation}
\bar{\eta}(0)_z=
\left\{
\begin{array}{ll}
1~~~{\mbox {if}}~~ z\in Q_x^i~~\forall i\\
\eta_z~~~{\mbox {otherwise}}
\end{array}
\right.
\end{equation}

The dynamics of the auxiliary process is chosen  as follows:

\begin{description}
\item (i)
The tagged particle can move from $x$ to $x+e_1$.
 The jump has rate  one if  $\eta\in{\cal{A}}_x^{+1}$ and $\bar{\eta}(0)\in {\cal{B}}_x^{+1}$ or $\eta\in{\cal{A}}_x^{+2}$
 and $\bar{\eta}(0){\cal{B}}_x^{+2}$, zero otherwise;
\item (ii)
The tagged particle can move from $x$ to $x-e_1$. The jump has rate  one if $\eta\in{\cal{A}}_x^{-1}$ and $\bar{\eta}(0)\in{\cal{B}}_x^{-1}$ or  $\eta\in{\cal{A}}_x^{-2}$ and $\bar{\eta}(0)\in{\cal{B}}_x^{-2}$, zero otherwise;

\item (iii)
The tagged particle can move from $x$ to $x+e_2$. The jump has rate one if $\eta\in{\cal{A}}_x^{+1}$ and $\bar{\eta}(0)\in {\cal{B}}_x^{+1}$ or $\eta\in{\cal{A}}_x^{-1}$ and $\bar{\eta}(0)\in{\cal{B}}_x^{-1}$, zero otherwise;

\item (iv)
The tagged particle can move from $x$ to $x-e_2$. The jump has rate one if $\eta\in{\cal{A}}_x^{-2}$ and $\bar\eta(0)\in{\cal{B}}_x^{-2}$ or $\eta\in{\cal{A}}_x^{+2}$ and $\bar{\eta}(0)\in {\cal{B}}_x^{+2}$, zero otherwise;

\item (v)
Configuration $\eta$ can be transformed in $\eta^{R_x^{+1},R_x^{-1}}$, namely the exchange of occupation variables
in $R_x^{+1}$ and $R_x^{-1}$ can be performed.
The move has rate one if $\eta\in{\cal{A}}_x^{+1}$ and  $\bar{\eta}(0)\in {\cal{B}}_x^{-1}$
 or $\eta\in{\cal{A}}_x^{-1}$ and $\bar{\eta}(0)\in {\cal{B}}_x^{+1}$, zero otherwise.

\item (vi)
Configuration $\eta$ can be transformed in $\eta^{R_x^{+2},R_x^{-2}}$, namely the
 exchange of occupation variables in $R_x^{+2}$ and $R_x^{-2}$ can be performed.
The move has rate one if $\eta\in{\cal{A}}_x^{+2}$ and $\bar{\eta}(0)\in {\cal{B}}_x^{-2}$
 or $\eta\in{\cal{A}}_x^{-2}$ and $\bar{\eta}(0)\in {\cal{B}}_x^{+2}$, zero otherwise.

\item (vii)
Configuration $\eta$ can be transformed in $\eta^{R_x^{+3},R_x^{-3}}$,
namely the exchange of occupation variables in $R_x^{+3}$ and $R_x^{-3}$ can be performed.
 The move has rate one if $\eta\in{\cal{A}}_x^{+1}$ and $\bar{\eta}(0)\in {\cal{B}}_x^{-3}$
or $\eta\in{\cal{A}}_x^{+2}$ and $\bar{\eta}(0)\in {\cal{B}}_x^{+3}$, zero otherwise;

\item (viii)
Configuration $\eta$ can be transformed in $\eta^{R_x^{+4},R_x^{-4}}$, namely the exchange of occupation variables
 in $R_x^{+4}$ and $R_x^{-4}$ can be performed.
 The move has rate one if $\eta\in{\cal{A}}_x^{-1}$
and $\bar{\eta}(0)\in {\cal{B}}_x^{-4}$ or $\eta\in{\cal{A}}_x^{-2}$ and
$\bar{\eta}(0) \in{\cal{B}}_x^{+4}$, zero otherwise;

\item (ix) All other moves are not allowed.
\end{description}
In the following we will show that the above choice of the rates is a suitable choice to perform the proof
of diffusivity, since the auxiliary
process has a positive diffusion coefficient and any move can be reconstructed by a finite
sequence of elementary moves allowed by KA.

Consider an initial configuration such that 
the tagged particle is
inside a frameable square of size $\Xi$ and such that all the sites in at least one of the sets $Q_{x(0)}^{(\pm 1)}, Q_{x(0)}^{(\pm 2)}$
 (see
figure \ref{RQ})
are empty , where $x(0)$ is the position of the tagged particle (i.e. $\eta(0)\in{\cal{A}}_x^i$ for some i).
Then, both
conditions will hold at any subsequent time.
Indeed, moves (i)--(iv) are such that the tagged particle
 remains always inside the empty two by two square. On the other hand, moves (v)--(viii)
 are devised in order that the only vacancies
that are moved during the process are those which belong at time zero to this two by two square, therefore sublattices
of size $\Xi$ that are frameable at time zero remain frameable at later times\footnote{ More precisely, sublattices
that are frameable in $\bar{\eta}(0)$ are frameable also for $\eta(t)$ at any t.}.
The fact that moves for the auxiliary process occur always inside frameable regions of size at most $\Xi$
implies that
 any move can be performed
through a finite sequence of elementary moves allowed by KA. Indeed,
by the properties of frameable configurations, any move inside a configuration of size $\Xi$
can be performed by a sequence of order $O(\Xi^2)$ moves with positive
rate for KA  dynamics \cite{KAprl,KAlong}.
By using path arguments analogous to those used in section \ref{triangular} for the triangular case, it is then possible to
establish inequality $D_S\geq c D_S^{aux}$, with
$c$ positive.
Let us shortly recall how this argument works and emphasize an important  difference occurring in this case.
For the triangular case we have defined an auxiliary process such that any move of the latter
can be performed by
a finite path of at most $n$ nearest neighbor moves allowed for the considered model.
Such path does not depend on the choice of the configuration.
Then, we have rewritten each term of  the
variational formula (\ref{variational}) of $D_S^{aux}$
as a telescopic sum on the exchanges along this path.
Finally, by using Cauchy Schwartz inequality,
 the fact that each possible move is used at most twice in the path and
by performing an exchange of variables, we concluded that
$D_S^{aux}\leq cD_S$ with $c>0$.
In this case we can proceed analogously, with the length $n$ of the paths at most $\Xi^2$.
However, the path now
depends on the whole configuration.
Indeed the sequence of allowed moves that one has to do in order to perform
a pair exchange depends on the position of the vacancies, see \cite{KAprl,KAlong}.
This yields in the inequality among $D_S^{aux}$ and $D_S$
an overall factor ${\cal N}=2^{\Xi^2}$ besides the factor due to $n=\Xi^2$.
 Let us explain in some detail this statement.\\
With a path argument analogous to the one done before,
we  rewrite each term in $D_S^{aux}$ corresponding to the exchange of particles in $R_x^i$, $R_x^{-i}$
as a telescopic sum over allowed exchanges for KA, namely

\begin{equation}
\label{telescopic3KA}
\left(f(\eta^{R_x^{+i},R_x^{-i}})-f(\eta)\right)^2
\leq \Xi^2\sum_{j=0}^n c_{x_{j-1},x_j}(\eta_j)\left(f(T_{x_{j-1},x_j}{\eta_j})-f(\eta_j)\right)^2
\end{equation}
where $c_{x,y}$ are the jump rate of KA model, $\eta_0=\eta,\eta_1,\dots,\eta_n=\eta^{R_x^{+i},R_x^{-i}}$
is the path of allowed elementary moves which connects $\eta$ to $\eta^{R_x^{+i},R_x^{-i}}$ and such
that $\eta_i=\eta_{i-1}^{x_{i-1},x_i}$ for a couple of nearest neighbors $\{x_i,x_{i-1}\}$. To obtain above inequality
 we
used Cauchy Schwartz inequality and the fact that $n\leq\Xi^2$.
In order to obtain from left and right hand side the terms which appear in the variational formula
 (\ref{variational}) for the
auxiliary process   and  for KA, respectively, we should average  inequality (\ref{telescopic3KA})
 over the Bernoulli measure conditioned to have a particle in zero.
As we already emphasized, the sequence $x_0\dots x_n$ of sites in which the exchange is
done,  depends on the positions of vacancies in configuration $\eta$.
Therefore, if we do the change of variable $\eta_i\to\eta$ in (\ref{telescopic3KA})
 and use the invariance of measure under exchange of variables,
many different terms on the left can give rise to the same term on the right.
Actually, the crucial thing to know is the following. To each configuration $\eta$
for which the exchange is allowed by the auxiliary process, associate the correspondent path $\eta_0,\dots,\eta_n$
in configuration space\footnote{ Of course there could be different sequences to perform the same move. However, one can
always  give a prescription associating one of them for any choice of $\eta$ and any give exchange.}.
 Then, for each elementary nearest neighbor exchange $e$, 
denote by ${\cal{N}}_e$ the number of different configurations
$\eta$ that use such exchange and
let ${\cal{N}}\equiv max_e~{\cal{N}}_e$.
Therefore ${\cal{N}}$ is the overall factor coming from possible overcounting of configuration when going
from the mean of the left hand side of (\ref{telescopic3KA})
to the terms in the variational formula (\ref{variational}) for $D_S$.
Physically ${\cal{N}}$ takes into account the most severe dynamical bottleneck
in phase space (see \cite{KAprl}).  
Moreover, since each path is composed of moves internal to the frameable region of size $\ell\leq \Xi$  which contains the tagged particle,
${\cal{N}}$ is for sure less or equal to the total number of configurations inside a square of size $\Xi$,
namely  inequality ${\cal{N}}\leq 2^{\Xi^2}$  holds.
Therefore,
we finally obtain, closely enough to unit density,

\begin{equation}
\label{daux}
D_S\geq c(1-\rho)^3~~\mu_{\rho,\Xi}({\cal{F}})~~\frac{1}{\Xi^4 2^{\Xi^2}}~~D^{aux}_S
\end{equation}
where $c$ is a positive constant.
The term $(1-\rho^3)\mu_{\rho,\Xi}({\cal{F}})$ comes from the condition that
 the configuration at time zero should have the tagged particle with three vacancies around  and be inside
a frameable square of size at most $\Xi$ and $D_S^{aux}$ is the diffusion coefficient of the auxiliary process
subject to this condition.
This ends the first part of the proof. In the following section  we shall show
that indeed $D_{S}^{aux}>0$.
\subsection{Lower bound for the self diffusion coefficient of the
auxiliary process}\label{lowerboundauxiliary}

Let us now prove that  $D^{aux}_S>0$, i.e. that diffusivity holds for the auxiliary process.
In this case, the mechanism which guarantees diffusivity is different
from the one discussed in section \ref{triangular}. 
Analogously to the KA on a triangular lattice (section \ref{triangular}), 
the auxiliary process we have just introduced is such that if the tagged particle is
at time zero in a two by two square of vacancies (i.e. $\eta(0)\in{\cal{A}}_{x(0)}^i$ for some i)
and inside a larger frameable square of size at most $\Xi$, both conditions will be always fulfilled
 at later times.
However, now it is not true that the tagged particle
 can always be moved in a chosen direction $e_i$ through a proper path.
For example, if we want to move it in direction $e_1$ this is possible only if
$\eta\in{\cal{A}}_{x}^{+1}$ or $\eta\in{\cal{A}}_{x}^{+2}$.
Otherwise, if $\eta\in{\cal{A}}_{x}^{-1}$
or $\eta\in{\cal{A}}_{x}^{-2}$,
the move is allowed only if before one makes the exchange $\eta\to\eta^{R_x^{+1},R_x{-1}}$ or
$\eta\to\eta^{R_x^{+2},R_x{-2}}$, respectively.
However these exchanges of rectangles (which are
the analogous of exchange $\eta\to\eta^{x-e_i, x+e_i}$ for RLG) are not always allowed.
Indeed (see rules (v)--(viii))
they have positive rate only if in the initial configuration the rectangular regions $R_{x}^{-1}$, respectively $R_{x}^{-2}$,
 do not contain vacancies
and are inside a frameable square of size at most $\Xi$.
Note that the rate of such exchanges (i.e. the rate of the exchange $\eta\to\eta^{R_x^{+1},R_x{-1}}$ and
$\eta\to\eta^{R_x^{+2},R_x{-2}}$ conditioned to the fact
that $\eta\in{\cal{A}}_{x}^{-1}$ or $\eta\in{\cal{A}}_{x}^{-2}$ respectively) does not depend
on the configuration $\eta$, but
is fixed once for all by the choice of the initial configuration $\eta(0)$. In other
 words
the choice of
the initial configuration fixes the {\sl good} rectangles that can be exchanged. This observation will allow us to map
the motion
of the two by two square of three vacancies plus tagged particle to a random
walk in a random  environment corresponding to the cluster of good rectangles.
We emphasize that this cluster does not change during dynamics, therefore the randomness of the environment
is {\sl quenched}.
Note that the probability $p_g$ for a given rectangle to be good
is greater than $\rho^2$ (the probability that both the sites inside the rectangle are occupied)
multiplied for the probability that it is inside a frameable region of size at most $\Xi$
which is almost one (this is the key physical ingredient). Therefore in the high density regime $p_g$
is  well above the threshold of conventional site percolation.
This implies that with unit probability the initial configuration has a percolating cluster of good
rectangles. By using that above the percolation threshold
random walk on random environment
has a positive diffusion coefficient \cite{DeMasiWick}
we will therefore obtain that the diffusion coefficient of
the auxiliary process is strictly positive.
In the following we will sketch the proof of the above argument in some detail.

Let $\eta_{(0,0)}=\eta(0)$ be the initial configuration.
Let us define the following sequence of configurations $\eta_{(m,n)}$ for $(m,n)\in\bZ^2$

\begin{equation}
\eta_{(m+1,n)}
\left\{
\begin{array}{ll}
\eta_{(m,n)}^{x,x+e_1}
~~~{\mbox {if}}~~ \eta_{(m,n)}\in {\cal{A}}^{+1}_{x},~~\bar{\eta}(0)
\in{\cal{B}}^{+1}_{x}\\
\phantom{caco}\\
\eta_{(m,n)}^{x,x+e_1}
~~~{\mbox {if}}~~ \eta_{(m,n)}\in {{\cal{A}}^{+2}_{x}},~~\bar{\eta}(0)
\in{\cal{B}}^{+2}_{x}\\
\phantom{caco}\\
\eta_{(m,n)}^{ R_{x}^{+1},R_{x}^{-1}}~~~{\mbox {if}}~~ \eta_{(m,n)}\in {\cal{A}}_{x}^{-1},~~
\bar{\eta}(0)\in {\cal{B}}^{+1}_{x}\\
\phantom{caco}\\
\eta_{(m,n)}^{ R_{x}^{+2}, R_{x}^{-2}}~~~{\mbox {if}}~~ \eta_{(m,n)}\in {\cal{A}}_{x}^{-2},~~\bar{\eta}(0)
\in{\cal{B}}^{+1}_{x}
\end{array}
\right.
\end{equation}

\begin{equation}
\eta_{(m,n+1)}
\left\{
\begin{array}{ll}
\eta_{(m,n)} ^{x,x+e_2}~~~{\mbox {if}}~~ \eta_{(m,n)}\in {\cal{A}}^{+1}_{x},~~
\bar{\eta}(0)\in{\cal{B}}^{+3}_{x}\\
\phantom{\eta_{(m,n)}~~~{\mbox {if}}~~ \eta_{(m,n)}\in
{\cal{D}}_2^x}\\
\eta_{(m,n)} ^{x,x+e_2}~~~{\mbox {if}}~~ \eta_{(m,n)}\in {\cal{A}}^{-1}_{x},~~
\bar{\eta}(0)\in{\cal{B}}^{+4}_{x}\\
\phantom{\eta_{(m,n)}~~~{\mbox {if}}~~ \eta_{(m,n)}\in
{\cal{D}}_2^x}\\
\eta_{(m,n)}^{ R_x^{+3},R_x^{-3}}~~~{\mbox {if}}~~ \eta_{(m,n)}\in {\cal{A}}_x^{+2},~~\bar{\eta}(0)\in{\cal{B}}^{+3}_{x}\\
\phantom{\eta_{(m,n)}~~~{\mbox {if}}~~ \eta_{(m,n)}\in {\cal{D}}_2^x}\\
\eta_{(m,n)}^{ R_x^{+4},R_x^{-4}}~~~{\mbox {if}}~~ \eta_{(m,n)}\in {\cal{A}}_x^{-2},~~\bar{\eta}(0)\in{\cal{B}}^{+4}_{x}
\end{array}
\right.
\end{equation}

\begin{equation}
\eta_{(m-1,n)}
\left\{
\begin{array}{ll}
\eta_{(m,n)}^{ x,x-e_1}~~~{\mbox {if}}~~ \eta_{(m,n)}\in {\cal{A}}^{-1}_{x},~~
\bar{\eta}(0)\in{\cal{B}}^{-1}_{x}\\
\phantom{\eta_{(m,n)}~~~{\mbox {if}}~~ \eta_{(m,n)}\in
{\cal{D}}_2^x}\\
\eta_{(m,n)}^{ x,x-e_1}~~~{\mbox {if}}~~ \eta_{(m,n)}\in {\cal{A}}^{-2}_{x},~~
\bar{\eta}(0)\in{\cal{B}}^{-2}_{x}\\
\phantom{\eta_{(m,n)}~~~{\mbox {if}}~~ \eta_{(m,n)}\in
{\cal{D}}_2^x}\\
\eta_{(m,n)}^{ R_x^{+1},R_x^{-1}}~~~{\mbox {if}}~~ \eta_{(m,n)}\in {\cal{A}}_x^{+1},~~\bar{\eta}(0)\in{\cal{B}}^{-1}\\
\phantom{\eta_{(m,n)}~~~{\mbox {if}}~~ \eta_{(m,n)}\in {\cal{D}}_2^x}\\
\eta_{(m,n)}^{ R_x^{+2},R_x^{-2}}~~~{\mbox {if}}~~ \eta_{(m,n)}\in {\cal{A}}_x^{+2},~~\bar{\eta}(0)\in{\cal{B}}^{-2}
\end{array}
\right.
\end{equation}

\begin{equation}
\eta^{(m,n-1)}
\left\{
\begin{array}{ll}
\eta_{(m,n)}^{ x,x-e_2}~~~{\mbox {if}}~~ \eta_{(m,n)}\in {\cal{A}}^{+2}_{x},~~
\bar{\eta}(0)\in{\cal{B}}^{-3}_{x}\\
\phantom{\eta_{(m,n)}~~~{\mbox {if}}~~ \eta_{(m,n)}\in
{\cal{D}}_2^x}\\
\eta_{(m,n)}^{ x,x-e_2}~~~{\mbox {if}}~~ \eta_{(m,n)}\in {\cal{A}}^{-2}_{x},~~
\bar{\eta}(0)\in{\cal{B}}^{-4}_{x}\\
\phantom{\eta_{(m,n)}~~~{\mbox {if}}~~ \eta_{(m,n)}\in {\cal{D}}_2^x}\\
\eta_{(m,n)}^{ R_x^{+3},R_x^{-3}}~~~{\mbox {if}}~~ \eta_{(m,n)}\in {\cal{A}}_x^{+1},~~\bar{\eta}(0)\in{\cal{B}}^{-3}\\
\phantom{\eta_{(m,n)}~~~{\mbox {if}}~~ \eta_{(m,n)}\in {\cal{D}}_2^x}\\
\eta_{(m,n)}^{ R_x^{+4},R_x^{-4}}~~~{\mbox {if}}~~ \eta_{(m,n)}\in {\cal{A}}_x^{-1},~~\bar{\eta}(0)\in{\cal{B}}^{-4}
\end{array}
\right.
\end{equation}
where $x=x(m,n)$ is the position
of the tagged particle in configuration $\eta_{(m,n)}$ (we drop the dependence of $x$ on $m$ and $n$ to get a more readable notation).
Note that given $\eta_{(m,n)}$ and $x(m,n)$ for a couple $(m,n)$,
using the above definitions one can reconstruct the whole sequence.\\
 Let us define a Markov process on $\bZ^2$ with generator $G$ acting on functions $f: (m,n)\to \bR$ as
\begin{eqnarray}
Gf(m,n)&:=& \sum_{i={\pm} 1}\lambda_{i}^1(m,n)\left(f(m+i,n)-f(m,n)\right)+
\nonumber\\
&&\sum_{i={\pm} 1}\lambda_{i}^2(m,n)\left(f(m,n+i)-f(m,n)\right)
\label{genG}
\end{eqnarray}
namely a two dimensional random walk with rates $\lambda_{{\pm} 1}^i$ for the jump in direction ${\pm} e_{i}$.
We can now chose these rates in order that
 $\eta_{(m(t),n(t))}=\eta(t)$ for the auxiliary process. More precisely, we chose the rates
in order that for any function $f(\eta)$ the expectation value over
the probability $\mu_t$ evoluted with the generator of the auxiliary process
coincides with the expectation over the measure on $(m(t),n(t))$
generated by
(\ref{genG}).
By considering the dynamics of the auxiliary process
and definition (\ref{genG}), one can directly check that the choice of  $\lambda_{{\pm} 1}^i$ which satisfies above requirement
is the following

\begin{eqnarray}
\label{randomrates}
\lambda_{+1}^{1}(m,n)&=&
\id_{ {\cal{A}}_{x}^{+1}}(\eta_{(m,n)})\id_{{\cal{B}}_x^{+1}\cap{\cal{B}}_x^{+2}}(\eta_{(0,0)})+\nonumber\\
&&\id_{ {\cal{A}}_{x}^{+2}}(\eta_{(m,n)})\id_{{\cal{B}}_x^{+1}\cap{\cal{B}}_x^{+2}}(\eta_{(0,0)})+\nonumber\\
&&\id_{ {\cal{A}}_{x}^{-1}}(\eta_{(m,n)})\id_{{\cal{B}}_{x}^{+1}}(\eta_{(0,0)})+\nonumber\\
&&\id_{ {\cal{A}}_{x}^{-2}}(\eta_{(m,n)})\id_{{\cal{B}}_{x}^{+2}}(\eta_{(0,0)})\nonumber\\
&&\\
\lambda_{-1}^{1}(m,n)&=&
\id_{ {\cal{A}}_{x}^{-1}}(\eta_{(m,n)})\id_{{\cal{B}}_x^{-1}\cap{\cal{B}}_x^{-2}}(\eta_{(0,0)})+\nonumber\\
&&\id_{ {\cal{A}}_{x}^{-2}}(\eta_{(m,n)})\id_{{\cal{B}}_x^{-1}\cap{\cal{B}}_x^{-2}}(\eta_{(0,0)})+\nonumber\\
&&\id_{ {\cal{A}}_{x}^{+1}}(\eta_{(m,n)})\id_{{\cal{B}}_{x}^{-1}}(\eta_{(0,0)})+\nonumber\\
&&\id_{ {\cal{A}}_{x}^{+2}}(\eta_{(m,n)})\id_{{\cal{B}}_{x}^{-2}}(\eta_{(0,0)})\nonumber\\
&&\\
\lambda_{+1}^{2}(m,n)&=&
\id_{ {\cal{A}}_{x}^2}(\eta_{(m,n)})\id_{{\cal{B}}_x^{+3}\cap{\cal{B}}_x^{+4}}(\eta_{(0,0)})+\nonumber\\
&&\id_{ {\cal{A}}_{x}^{-1}}(\eta_{(m,n)})\id_{{\cal{B}}_x^{+3}\cap{\cal{B}}_x^{+4}}(\eta_{(0,0)})+\nonumber\\
&&\id_{ {\cal{A}}_{x}^{2}}(\eta_{(m,n)})\id_{{\cal{B}}_{x}^{+3}}(\eta_{(0,0)})+\nonumber\\
&&\id_{ {\cal{A}}_{x}^{-2}}(\eta_{(m,n)})\id_{{\cal{B}}_{x}^{+4}}(\eta_{(0,0)})\nonumber\\
&&\\
\lambda_{-1}^{2}(m,n)&=&
\id_{ {\cal{A}}_{x}^{2}}(\eta_{(m,n)})\id_{{\cal{B}}_x^{-3}\cap{\cal{B}}_x^{-4}}(\eta_{(0,0)})+\nonumber\\
&&\id_{ {\cal{A}}_{x}^{-2}}(\eta_{(m,n)})\id_{{\cal{B}}_x^{-3}\cap{\cal{B}}_x^{-4}}(\eta_{(0,0)})+\nonumber\\
&&\id_{ {\cal{A}}_{x}^{+1}}(\eta_{(m,n)})\id_{{\cal{B}}_{x}^{-3}}(\eta_{(0,0)})\nonumber\\
&&\id_{ {\cal{A}}_{x}^{-1}}(\eta_{(m,n)})\id_{{\cal{B}}_{x}^{-4}}(\eta_{(0,0)})
\end{eqnarray}
Let us define also

\begin{eqnarray}
\label{randomrates2}
\bar\lambda_{+1}^{1}(m,n)&=&
\id_{{\cal{B}}_{x}^{+1}\cap {\cal{B}}_{x}^{+2}}(\eta_{(0,0)})\nonumber\\
\bar\lambda_{-1}^{1}(m,n)&=&
\id_{{\cal{B}}_{x}^{-1}\cap{\cal{B}}_{x}^{-2}}(\eta_{(0,0)})\nonumber\\
\bar\lambda_{+1}^{1}(m,n)&=&
\id_{{\cal{B}}_{x}^{+3}\cap {\cal{B}}_{x}^{+4}}(\eta_{(0,0)})\nonumber\\
\bar\lambda_{+1}^{1}(m,n)&=&
\id_{{\cal{B}}_{x}^{-3}\cap {\cal{B}}_{x}^{-4}}(\eta_{(0,0)})\nonumber\\
&&
\end{eqnarray}
and

\begin{eqnarray}
\bar G f(m,n)&:=& \sum_{i={\pm} 1}\bar\lambda_{i}^1(m,n)\left(f(m+i,n)-f(m,n)\right)+
\nonumber\\
&&\sum_{i={\pm} 1}\bar \lambda_{i}^2(m,n)\left(f(m,n+i)-f(m,n)\right)
\label{genG2}
\end{eqnarray}
>From definitions (\ref{randomrates}) and (\ref{randomrates2}) it is
immediate to check that $\lambda_{{\pm} 1}^i\geq\bar\lambda_{{\pm}
1}^i$. Because the variational formula (\ref{variational}) implies
that the self diffusion coefficient is a monotonic increasing function
of the jump rates we find that $D_{S}$ for the random walk process with
rates $\lambda_{{\pm} 1}^i$ is certainly larger than the one for the
process with rates $\bar\lambda_{{\pm}
1}^i$.
By recalling definition (\ref{cB}) and results on crossover length, we know that
 for sufficiently high density the probability w.r.t. Bernoulli measure of events

$${\cal{B}}_{x}^{+1}\cap {\cal{B}}_{x}^{+2},$$
$${\cal{B}}_{x}^{-1}\cap {\cal{B}}_{x}^{-2},$$
$${\cal{B}}_{x}^{+3}\cap {\cal{B}}_{x}^{+4},$$
$${\cal{B}}_{x}^{-3}\cap {\cal{B}}_{x}^{-4},$$
is almost one, and therefore greater then threshold probability for conventional percolation on the square lattice.
 Hence, by using the result in \cite{DeMasiWick} which establishes a central limit
 theorem for random walk in random environment when bond probability
is greater than percolation threshold we find that $D_{S}$ for the 
process with rates $\bar\lambda_{{\pm}
1}^i$ (and therefore $D_{S}$ for the 
process with rates $\lambda_{{\pm}
1}^i$) is strictly positive.\footnote{
Note that the percolation problem we consider is a site percolation
problem in which the site probability is correlated over a { \it finite}
distance equal to $2\Xi(\rho)$ whereas in \cite{DeMasiWick} it was analyzed
only the case with independent bond probability. However we expect
that their result can be generalized to our case. Furthermore
from the physical point of view there is no doubt that
$D_{S}$ will be positive in our case of correlated site disorder 
whenever a giant cluster exists (except at the critical point).} 
Moreover, since when $m$ goes to $m+2$ ($n$ to $n+2$) the first (second) coordinate of the tagged particle position increases at least of one unit, inequality

\begin{equation}
\mu_t\left(x_1(t)^2+x_2(t)^2\right)\geq \frac{1}{4}E\left(m(t)^2+n(t)^2\right)>ct
\end{equation}
holds, where $\mu_t$ is the evoluted of initial measure $\mu_{\rho,0}$ under the
auxiliary process and $c$ a positive constant.
This allows us to conclude that $D^{aux}_S>0$ at any $\rho<1$ which, together with inequality (\ref{daux}),
implies $D_S>0$ for KA model.

\section{Conclusion}\label{conclusion}
In this work we have presented a general procedure, focusing on the KA
model as an example, that allows one to
prove lower and upper bounds on interacting particle systems on a
lattice in the case of vanishing rates. 
This is a generalization of the Spohn's proof \cite{Spohn,Sart}
for RLG. In particular, focusing on KA $s=1$ model on a
triangular lattice as an example, we show how to obtain the exact density dependence of $D_{S}$
in the high density limit for KCLG with finite size defects (see also
\cite{BT}).
Whereas for highly cooperative KCLG, i.e. when the size
of the defects diverges approaching unit density, our procedure allows to prove
diffusivity, $D_{S}>0$, at any density smaller than one as we have
shown for the KA $s=1$ model on a square lattice.\\
As we have stressed previously our method is completely general
and can be applied to general (short-range) interacting particle
systems on a lattice, including systems with a non--trivial equilibrium
measure as for example the statically constrained models introduced in
\cite{BiroliMezard}.
In order to be successful, it just needs one crucial key physical
property. One has to know that, for a random equilibrium configuration, 
all the particle exchanges inside a finite box
around a fixed site, say the origin, can be performed with a very high
probability $p$ through a suitable path in
configuration space (allowed by the kinetic rules) 
that involves particles at most at distance $\Xi$
from the origin. Furthermore one needs also that $p$ can be chosen arbitrary close
to one taking a suitably large $\Xi$. \\
This property is certainly valid if, given two equilibrium
configurations taken at random inside a box of size $L$,
one can show that there exists with probability $p'$ at least a path in
configuration space, allowed by the kinetic rules, that connects the
two configurations and that $\lim_{L\rightarrow \infty}{p'}=1$
\footnote{The boundary condition for the box has to be chosen such
that it is the worst possible in order to find the path connecting
two configurations. For example for the KA one would choose boundary
conditions that are equivalent to embed the box in a completely filled lattice.
}.
Note that is just equivalent to say that the system at hand is
irreducible in the thermodynamic limit. More correctly, in the
thermodynamic limit one irreducible component $\Omega $ covers all the
configuration space, i.e. an equilibrium configuration taken at random
belongs to $\Omega $ with probability one.
Therefore our results, combined with the ones obtained for Brownian
interacting particle systems \cite{BrownianDS}, strongly suggest that the only case in which a
dynamical arrest (at which $D_{S}$ vanishes at a finite
temperature/chemical potential) might happen is only when 
a irreducible-reducible transition takes place and that
one has to identify the two types of transition. \\
To our knowledge, none of the
short-range interacting particle systems which have been considered 
so far (on a lattice as well as in the continuum) has been proved to
display such reducibility transition in dimension larger than one.

Finally, we want to stress again that the non-vanishing of the
self-diffusion coefficient $D_{S}$ does not imply that the structural 
relaxation time scale $\tau_{\alpha }$ cannot diverge. In particular, as discussed
previously, at a second order phase transition the structural
relaxation time scale diverges whereas the self-diffusion coefficient stays finite.
Thus, the decoupling between $D_{S}$ and $1/\tau_{\alpha }$ is {\it a
necessary condition for the existence of an ideal glass transition}
taking place at finite temperature and chemical potential. In experiments
on fragile liquids \cite{Decoupling} a decoupling is indeed observed between $D_{S}$ and
$1/\tau_{\alpha }$ (more correctly between $D_{S}$ and the viscosity
$\eta$) but not as strong as one would have very close to a phase
transition, i.e. $D_{S}\tau_{\alpha }\propto \tau_{\alpha }$.
\\
\\
It is pleasure to thank Daniel S. Fisher for all the discussions that we
had on the subject of this work and, more generally, on kinetically
constrained models for glasses.

\addcontentsline{toc}{chapter}{{\bfseries Bibliography}}

\end{document}